\documentclass[10pt]{ip-journal}

\usepackage{graphicx}
\begin{document}

\title[Enlarged Transformation Group: Applications]{Enlarged Transformation Group: \\ Star Models,  Dark Matter Halos  \\ and  Solar System Dynamics }

\author{Edward Lee Green}

\address{University of North Georgia, Dahlonega, GA 30597}
\email{egreen@ung.edu}
\begin{abstract}
Previously a theory has been presented which extends the geometrical structure of a real four-dimensional space-time via a field of orthonormal tetrads with an enlarged transformation group.  This new transformation group, called the conservation group, contains the group of diffeomorphisms as a proper subgroup and we hypothesize that it is the foundational group for quantum geometry.  The fundamental geometric object of the new geometry is the curvature vector, $C_\mu$.  Using the scalar Lagrangian density $C^\mu C_\mu \sqrt{-g}\,$, field equations for the free field have been obtained which are invariant under the conservation group.  In this paper, this theory is further extended by development of a suitable Lagrangian for a field with sources. Spherically symmetric solutions for both the free field and the field with sources are given.  A stellar model and an external, free-field model are developed. The theory implies that the external stress-energy tensor has non-compact support and hence may give the geometrical foundation for dark matter.  The resulting models are compared to the internal and external Schwarzschild models.  The theory may explain the Pioneer anomaly and the corona heating problem. (PACS 04.50.-h, 12.10.-g,04.40.-b)
\end{abstract}

\maketitle

\section{Introduction}

Let ${X}^4$ be a 4-dimensional space with orthonormal tetrad $h^{i}_{\,\,\mu}$.  Then a metric
$g_{\mu \nu}$ may be defined on ${ X}^4$ by $g_{\mu\nu}=\eta_{ij}\, h^i_{\,\,\mu}h^j_{\,\,\nu}$ where $\eta_{ij} =
diag\bigl\{-1,1,1,1\bigr\}$.  Whereas Einstein extended special relativity to general relativity by extending the group of transformations from the Lorentz group to the group of diffeomorphisms, Einstein later suggested that that a unified field theory may be obtained by extending the diffeomorphisms to a larger group [1].  Einstein was also led by the principle that the speed of light was constant.  Consistent with Einstein's approach, we look for the largest group of transformations for which the wave equation, $\Psi^\alpha_{\;\; ; \alpha} = 0$, is covariant.   This is the guiding principle for our theory.

Let $\tilde{V}^\alpha$ be a vector density of weight $+1$.  Then a conservation law of the form $\tilde{V}^\alpha_{\;\; , \alpha}=0$ is invariant under all transformations satisfying
\begin{equation}x^\nu_{\;\; ,\overline{\alpha}}\bigl(x^{\overline{\alpha}}_{\;\; ,\nu ,\mu} -
x^{\overline{\alpha}}_{\;\; ,\mu , \nu} \bigr) = 0 \quad .  \end{equation}
This property defines the group of conservative transformations, of which, the group of diffeomorphisms is a proper subgroup [2].  Since the wave equation may be written as $\tilde{V}^\alpha_{\;\; ,\alpha} = 0 $ with $\tilde{V}^\alpha = \sqrt{- g\,}\, \Psi^\alpha$, we see that the conservation group is "the largest group of coordinate transformations under which the equation for the propagation of light is covariant" [3].  The conservation group shows potential for being the fundamental group for a unified field theory, a theory encompassing all the forces of nature [2-6].

The argument that accelerated observers should be on equal footing led Einstein to general relativity.  We have argued that requiring that quantum observers be on equal footing leads to the conservation group [3].  If we, as observers consider ourselves to be classical (non-quantum) observers, we will have a preference for the manifold view for what we observe. In truth, we are are quantum observers and hence some "fuzziness" in our observations as well as the observations of other observers is present.
Suppose $x^\mu$ are used as coordinates on a neighborhood of "our manifold" and $x^{\bar{\mu}}$ are used as coordinates on a neighborhood of a second observer.   If $x^{\bar{\mu}}_{\; ,\nu}$ is non-diffeomorphic but is conservative, satisfying (1), then $x^{\bar{\mu}}$ may be interpreted as anholonomic coordinates for "our manifold"[7].  Alternatively, the transformation from $x^\mu$ to $x^{\bar{\mu}}$ may be viewed as a transformation from one manifold to a second manifold.  This second manifold has a different metric and a different curvature tensor $R^\alpha_{\;\;\beta\mu\nu}$.  When manifolds ${M}_1$ and ${M}_2$ are related by such a conservative (but, possibly non-diffeomorphic) transformation, we say they are in the same quantum family of manifolds denoted by ${M}_1 \sim {M}_2$.    There are calculations (some in this paper) that suggest that the curvature vector given below may be related to the mass of a classical particle.  We see that quantum observers related by (1) agree on the speed of light and, if these hints are correct, on the value of the masses of classical particles present (if any).

The conservative group property (1) ensures that these quantum observers, one using manifold ${M}_1$ and the other using manifold ${M}_2$ are on equal footing.  We know from experience, however, that the classical solution is preferred and hence there is a preferred manifold.  That preferred manifold is the classical manifold, ${M}_0$ with curvature tensor $R^\alpha_{\;\beta\mu\nu}$.  A probability amplitude, constructed from the curvature tensor and/or appropriate contractions may turn out to be the correct probability amplitude for the quantum geometry.  We will give a tentative expression below for this probability amplitude.  In the sum over all possible manifolds (analogous to the path integral sum over histories), the classical manifold receives preference - nonclassical solutions tend to cancel out.

The neighborhood of the second observer continues to make geometric sense to the first observer, but only at the infinitesimal level.  We see that neighborhoods upon which the coordinate systems of the second observer make sense to us as the first observer have shrunk from global (special relativity) to local (general relativity) to infinitesimal (conservation group theory).  We stipulate that we may begin the setup of our theory by defining $h^i_{\;\alpha}$ as a function of $x^\mu$ so that the corresponding metric correctly models the gravitational fields on the boundary of a region.  This will give a set of admissible manifolds.  Then we may determine the preferred classical geometry as the manifold ${M}_0$.   The full quantum geometry, ${Q}$, is associated with the family of manifolds related to ${M}_0$ via  conservative transformations, i.e. ${Q} =\{ {M} \big| {M} \sim {M } _0 \}$.

If a transformation from $x^\mu$ to $x^{\bar{\mu}}$ is conservative, but not diffeomorphic, then, in addition to changing the curvature, this transformation will cause expressions such as  $[\partial_\mu ,\partial_\nu ]f=0$ to be nonzero in the new space:       $[\partial_{\bar{\mu}}, \partial_{\bar{\nu}}]f \neq 0$.  However, since we are requiring the transformation to be in the group of conservative transformations, we are not simply abandoning the diffeomorphism condition in an {\it ad hoc} manner.

In an effort to model dark matter cosmic acceleration, many theorists have simply modified general relativity in some fashion.  We claim that our modification which is based on an enlargement of  the transformation group is perhaps the only one with a solid guiding principle.  Einstein himself felt it was a mistake to simply add a cosmological constant $\Lambda$.  Recently, theoretical developments of $f(R)$ gravity [8], quintessence [9] and other modifications of general relativity [10] have a similar {\it ad hoc} flavor.

The geometrical content of the theory based on the conservation group is determined by $C_\alpha \equiv h_i^{\,\,\nu}\bigl(h^i_{\,\, \alpha ,\nu}-h^i_{\,\, \nu ,\alpha} \bigr) \; = \gamma^\mu_{\;\;\alpha\mu}$, where the Ricci rotation coefficient is given
by $\gamma^i_{\;\;\mu\nu}=h^i_{\;\mu ;\nu}$ [2-6]. Pandres calls $C_\alpha$ the curvature vector. He shows that $C_\alpha$ is covariant under transformations from $x^\mu$ to $x^{\overline{\mu}}$ if and only if the transformation is conservative and thus satisfies (1).  A suitable scalar Lagrangian for the free field is given by
\begin{equation} {L}_f= \frac1{16\pi}\int C^\alpha C_\alpha \, h \; d^4 x  \end{equation} where $h = \sqrt{-g}$ is the determinant of the tetrad.

Using $h^i_{\, \mu}=h^I_{\, \mu}\Lambda^i_I$, we have extended the field variables [5] to include the tetrad
$h^I_{\;\mu}$ and 4 internal vectors $\Lambda^i_I$, with internal space variable $x^I$. The distinctive feature of the internal space is that its metric is Lorentzian, i.e., $g_{IJ}= \eta_{IJ}\equiv diag(-1,1,1,1)$.  With this extension, the covariant derivative has been extended to be invariant under a larger group of transformations on $x^I$ as well as $x^\mu\;$ [5].    The definition of the Ricci rotation coefficient is also extended using the $\Lambda^i_I$ to
\begin{equation}\Upsilon^\alpha_{\;\;\mu\nu}\equiv h_I^{\;\alpha}h^I_{\;\mu
;\nu}+h_i^{\;\alpha}h^I_{\;\mu}\Lambda^i_{I,\nu} \end{equation} and the definition of $C_\alpha$ is
also extended to $C_\alpha \equiv \Upsilon^\mu_{\;\;\alpha\mu}$. Using these extended
Ricci rotation coefficients, one finds that
\begin{equation}C^\alpha C_\alpha = R + \Upsilon^{\alpha \beta \nu} \Upsilon_{\alpha \nu \beta}-2C^\alpha_{\; ;\alpha}
- \eta^{ij}h_j^{\;\nu}h_I^{\;\alpha}(\Lambda^I_{i,\alpha ,\nu}-\Lambda^I_{i,\nu
,\alpha})\quad , \end{equation}
where $R$ is the usual Ricci scalar curvature.  Comparing with GR we see that the Lagrangian density of the free field contains additional terms.  These terms correspond to quantum corrections to our manifold (classical) interpretation of physical space [4,6].

The motion of a free particle or photon in the inertial coordinate system is given by \begin{equation} \frac{d^2 x^i}{ds^2}=0\, , \end{equation} where $-ds^2=\eta_{ij}dx^i dx^j$.  This equation when transformed to internal coordinates, $x^I$ is
\begin{equation} \frac{d^2x^I}{ds^2}=-\Lambda^I_i \Lambda^i_{J,K} \frac{dx^J}{ds\;}\frac{dx^K}{ds\;} \; , \end{equation}
where the right hand side of this equation is zero when there are no internal forces.  Since $\eta_{IJ}$ corresponds to the flat metric, we naturally interpret the right hand side of (6) as a force.   The $\Lambda^i_I$ are thus internal fields that via $\Lambda^i_{I,J}$ correspond to electroweak and strong interactions.  In the manifold view, with coordinates $x^\alpha$  equation (6) becomes
\begin{equation} \frac{d^2 x^\alpha}{ds^2}+
\Gamma^\alpha_{\mu\nu}\frac{dx^\mu}{ds\;}\frac{dx^\nu}{ds\;} = -
\Upsilon^\alpha_{\;\;\mu\nu}\frac{dx^\mu}{ds\;}\frac{dx^\nu}{ds\;} \; . \end{equation} The right hand is partly generated from the internal forces since from (3) one sees that this equation of motion depends on $\Lambda^i_{I,\nu}$.

Setting the variations of ${L}_f$ with respect to $h^I_{\;\;\mu}$ and $\Lambda^i_I$ equal to zero along with the assumption that we may always choose $\Lambda^i_I$ to correspond to a complex Lorentz transformation (since $h^i_{\;\mu}=h^I_{\;\mu}\Lambda^i_I$), yields the field equations [5]
\begin{equation}  C_\mu = 0 \quad .  \end{equation}

One feature of the extended theory with field variables $h^I_{\;\mu}$ and $\Lambda^i_I$ is that the internal fields associated with $\Lambda^i_I$ may be specified after finding a tetrad $h^I_{\;\alpha}$ which satisfies the condition $h_I^{\;\nu}\bigl(h^I_{\; \mu ,\nu}- h^I_{\; \nu , \mu}\bigr)=0$.  This tetrad $h^I_{\;\alpha}$ yields a Riemannian manifold with corresponding metric, $g_{\mu\nu} = \eta_{IJ}h^I_{\;\mu}h^J_{\;\nu}$.  Changes in $\Lambda^i_I$ have no effect on this manifold [5].
{\it Since this paper is concerned with gravitational implications of the the theory we will assume for the remainder of this paper that we are working with a solution of the field equations for which $\Lambda^I_i =\delta^I_i $ (i.e., no internal fields). Thus $h^i_{\; \mu}=h^I_{\;\mu}\delta^i_I$, i.e., the matrices for $h^i_{\;\mu}$ and $h^I_{\;\mu}$ are the identical.}  In this case, an identity for  the Einstein tensor is
\begin{eqnarray}  G_{\mu\nu}= & C_{\mu ; \nu}- C_\alpha \Upsilon^\alpha_{\; \mu\nu} -g_{\mu\nu}C^\alpha_{\; ;\alpha}-\frac12 g_{\mu\nu}C^\alpha C_\alpha   \nonumber \\ \quad & +\Upsilon^{\;\;\alpha}_{\mu \;\; \nu ;\alpha}+\Upsilon^\alpha_{\;\;\sigma
\nu} \Upsilon^\sigma_{\;\; \mu \alpha} + \frac12 g_{\mu\nu}\Upsilon^{\alpha \beta
\sigma} \Upsilon_{\alpha \sigma \beta} \nonumber \end{eqnarray}  This expression is not manifestly symmetric in $\mu$ and $\nu$, but the left-hand side is symmetric in its lower indices and hence the right-hand side must be as well.  Thus we use a symmetrized expression to ensure this.  Define for general $K_{\mu\nu}$, the symmetrized tensor by $K_{(\mu\nu)}=\frac12(K_{\mu\nu}+K_{\nu\mu})$.   Using (8) we see that the field equations may be also expressed in the form
\begin{equation}G_{\mu\nu} = \Upsilon^{\;\;\alpha}_{(\mu \;\; \nu) ;\alpha}+  \Upsilon^\alpha_{\;\;\sigma (\nu} \Upsilon^\sigma_{\;\; \mu) \alpha}  + \frac12 g_{\mu\nu}\Upsilon^{\alpha \beta
\sigma} \Upsilon_{\alpha \sigma \beta}\quad \equiv 8\pi \bigl(\mathbf{T}_{\rm f}\bigr)_{\mu\nu}  \end{equation}
with free field stress energy tensor $\mathbf{T}_{\rm f}$. The terms of $\mathbf{T}_{\rm f}$ suggest that this new geometry produces a stress energy tensor with additional terms that could be the stress energy tensor for dark matter or dark energy [6].

In the presence of sources the Lagrangian is of the form
\begin{equation} {L}= {L}_{\rm f}+{L}_{\rm s} =  \int \biggl( \frac1{16\pi}C^\alpha C_\alpha + L_s \biggr) \, h \; d^4 x  \end{equation}  where $L_s$ ( a function of $h^i_{\;\mu}\;$) is the appropriate Lagrangian density function for the source.
In this case $C_\alpha$ is nonzero  and variation of (10) with respect to the tetrad results in
$$\int \Biggl[\frac{1}{16\pi}\biggl(C_{(\mu ;\nu)}- C_\alpha \Upsilon^\alpha_{\; (\mu\nu)} -\frac12 g_{\mu\nu}C^\alpha C_\alpha - g_{\mu\nu}C^\alpha_{\; ;\alpha} \biggr) -\frac12 (T_{\rm s})_{\mu\nu}\Biggr] h\, h^{i\nu} \delta h_i^{\;\mu} \, d^4 x = 0$$
Here, $(T_{\rm s})_{\mu\nu}$ is the usual stress-energy tensor of the source for the standard theory [11].  Thus
\begin{equation} C_{(\mu ;\nu)}- C_\alpha \Upsilon^\alpha_{\; (\mu\nu)} -\frac12 g_{\mu\nu}C^\alpha C_\alpha - g_{\mu\nu}C^\alpha_{\; ;\alpha} = 8\pi (T_{\rm s})_{\mu\nu} \end{equation}
and also we have the following identity for the Einstein tensor,
\begin{equation} G_{\mu\nu}=  \biggl( \Upsilon^{\;\;\alpha}_{(\mu \;\; \nu ) ;\alpha}+\Upsilon^\alpha_{\;\;\sigma
 (\nu} \Upsilon^\sigma_{\;\; \mu ) \alpha} + \frac12 g_{\mu\nu}\Upsilon^{\alpha \beta
\sigma} \Upsilon_{\alpha \sigma \beta} \biggr) + 8\pi (T_{\rm s})_{\mu\nu}  \quad   \end{equation}
or
\begin{equation} G_{\mu\nu} =  8\pi\bigl(T_{\rm f}\bigr)_{\mu\nu}  + 8\pi\bigl(T_{\rm s}\bigr)_{\mu\nu}  \qquad .  \end{equation}  We call $\mathbf{T}_{\rm f}$ the free field stress energy and $\mathbf{T}_{\rm s}$ the stress energy for the source.

\par \phantom{D} \par

\section{Spherically symmetric solutions.}

\subsection{ Free Fields.}

We now exhibit spherically symmetric solutions of the field equations for a free field (5).  Let $r=\sqrt{(x^1)^2+(x^2)^2+(x^3)^2}$.  If $f(r)$ is a positive differentiable function of $r$,
then the tetrad field given by
\begin{equation} h^i_{\;\; \mu} = \delta^i_0 \delta^0_\mu \sqrt{f(r)} + \frac1{\root 4 \of{f(r)}} (\delta^i_1
\delta^1_\mu + \delta^i_2 \delta^2_\mu + \delta^i_3 \delta^3_\mu ) \end{equation} yields
$C_\mu=0$ and hence is a solution of the field equations (5).  The line element (metric) in
spherical coordinates  is given by
\begin{equation} ds^2 = -f(r)dt^2 + \frac1{\sqrt{f(r)}} dr^2 + \frac{r^2}{\sqrt{f(r)}} d\theta^2 +
\frac{r^2\sin^2\theta}{\sqrt{f(r)}} d\phi^2 \quad .  \end{equation} This is the line element (metric) in isotropic spherical coordinates.  Now change the radial
coordinate $r \to \overline{r}$ so that $\overline{r}^2 = \frac{r^2}{\sqrt{f(r)}}$ and
$f(r) = e^{2\Phi(\overline{r})}$.  Since these are differentiable functions, this
change of coordinates $(t,r,\theta,\phi) \to (t,\overline{r},\theta,\phi)$ is a
diffeomorphism and hence the field equations remain satisfied.  The mapping $r \to
\overline{r}$ is the simply the inverse of the function
$r=r(\overline{r})=\overline{r}e^{\frac12 \Phi(\overline{r})}$.  After this change in the
radial coordinate $r$, we will now rename $\overline{r}$ as simply $r$.  The tetrad in
spherical coordinates may be expressed by
\begin{equation} h^i_{\;\; \mu} = \left[  \begin{array}{cccc}
\; e^{\Phi} & 0 & 0 & 0 \\
0 &\; \bigl(1+\frac12 r\Phi^\prime \bigr)\sin \theta \cos \phi \; & \; r\cos\theta\cos\phi \; & \; -r\sin\theta\sin\phi \\
0 & \bigl(1+\frac12 r\Phi^\prime \bigr)\sin\theta\sin\phi & r\cos\theta\sin\phi & \;\; r\sin\theta\cos\phi \\
0 & \bigl(1+\frac12 r\Phi^\prime \bigr)\cos\theta\qquad & -r\sin\theta\qquad & 0
\end{array}
\right]
\end{equation}   where the upper index refers to the row and the prime indicates differentiation with respect to $r$.  One finds that
$C_\mu=0$ for this tetrad.   The new metric is
\begin{equation}ds^2= -e^{2\Phi(r)} dt^2 +  \bigl(1+\frac12 r\Phi^\prime(r) \bigr)^2dr^2+r^2d\theta^2
+r^2\sin^2\theta d\phi^2 \quad . \end{equation} After a long, but straightforward
calculation, one finds that the Einstein tensor equals a diagonal tensor which is in general nonzero: $G_{\mu\nu}=8\pi \bigl(T_{\rm f}\bigr)_{\mu\nu}$.  The non-zero
components are (with $\Phi$ representing $\Phi(r)$)
\begin{equation}G_{tt} =\,8\pi \bigl(T_{\rm f}\bigr)_{tt}\;=\; \frac{e^{2\Phi}\biggl(
\frac18(r\Phi^\prime)^3+\frac34(r\Phi^\prime)^2+2r\Phi^\prime+r^2\Phi^{\prime\prime}\biggr)
}{r^2\biggl(1+\frac12r\Phi^\prime\biggr)^3} \quad ,  \end{equation}
\begin{equation}G_{rr}=\; 8\pi \bigl(T_{\rm f}\bigr)_{rr} \,=\; \frac{r\Phi^\prime - \frac14(r\Phi^\prime)^2}{r^2}
\end{equation} and
\begin{eqnarray} \frac{G_{\theta\theta}}{r^2}=&\frac{8\pi T_{\theta\theta}}{r^2}=& \frac{\frac12(r\Phi^\prime)^3+(r\Phi^\prime)^2
+\frac12 r\Phi^\prime+\frac12r^2\Phi^{\prime\prime}}{r^2\bigl(1+\frac12r\Phi^\prime \bigr)^3} \\
\frac{G_{\phi\phi}}{r^2\sin^2\theta} = &\frac{8\pi T_{\phi\phi}}{r^2\sin^2\theta}
=& \frac{\frac12(r\Phi^\prime)^3+(r\Phi^\prime)^2
+\frac12 r\Phi^\prime+\frac12r^2\Phi^{\prime\prime}}{r^2\bigl(1+\frac12r\Phi^\prime \bigr)^3} \quad . \nonumber
\end{eqnarray}
One difference between this and the Schwarzschild metric [12] is that there is
only one unknown function ($\Phi(r)$) instead of two (the standard $\Lambda(r)$ and
$\Phi(r)$ functions).

We will first work on the $G_{tt}$ term.  One finds that
\begin{equation}e^{-2\Phi(r)}G_{tt}=\frac{2}{r^2} \cdot \frac{d\;}{dr}\biggl( \frac{r}2 -
\frac{r}{2(1+\frac12r\Phi^\prime)^2} \biggr) \equiv \frac2{r^2} w^\prime(r)\equiv 8\pi \rho_f \quad , \end{equation}
where $w(r)\equiv \frac{r}2 - \frac{r}{2(1+\frac12 r\Phi^\prime)^2}$.
Hence \begin{equation}\Phi^\prime(r)= \frac2r\biggl[\bigl(1-\frac{2w(r)}{r}\bigr)^{-\frac12} - 1\biggr] \quad . \end{equation}
Thus
\begin{equation}g_{rr}=\bigl(1+\frac12r\Phi^\prime\bigr)^2=\biggl(1-\frac{2w(r)}{r}\biggr)^{-1} \; \; , \end{equation} and
\begin{equation}g_{tt}= -e^{2\Phi(r)} \quad \mathtt{ , where } \;\;   \Phi(r)= \int \frac2r\biggl[\Bigl(1-\frac{2w(r)}{r}\Bigr)^{-\frac12} - 1\biggr] \, dr \quad \end{equation}  (this defines $\Phi(r)$ up to a constant).  The function $w(r)$ (as shown below) is related to the mass inside a ball of radius $r$ for the free field and $\rho_f$ represents the density of the free field in the manifold interpretation.

Let $p_R$ represent the radial pressure of the free field.  Then one finds [12] that the radial pressure of the free field is given by
\begin{equation}8\pi p_R= \frac{G_{rr}}{\bigl(1+\frac12r\Phi^\prime\bigr)^2} =
\frac{r\Phi^\prime - \frac14(r\Phi^\prime)^2}{r^2\bigl(1+\frac12r\Phi^\prime\bigr)^2} \end{equation} and from (22) one finds that
\begin{equation}8\pi p_R = \frac{ 4r\sqrt{1-\frac{2 w(r)}{r}} - 4r + 6w(r)}{r^3} \quad . \end{equation}

Let the tangential pressure of the free field be denoted by $p_T$.  We also find that $8\pi
p_T=\frac{G_{\theta\theta}}{r^2}=\frac{G_{\phi\phi}}{r^2\sin^2\theta}$ and thus,
\begin{equation}8\pi p_T=\frac{\frac12(r\Phi^\prime)^3+(r\Phi^\prime)^2
+\frac12r\Phi^\prime+\frac12r^2\Phi^{\prime\prime}}{r^2\bigl(1+\frac12r\Phi^\prime \bigr)^3} \quad .
 \end{equation} Using (22), the tangential pressure may be expressed in terms of $w(r)$ and $r$ by
\begin{equation}8\pi p_T=\frac{8r-9w(r)-8r\sqrt{1-\frac{2w(r)}{r}}+r w^\prime(r)}{r^3} \quad .
\end{equation}
Since $p_R\neq p_T$ there are shear stresses and we see that $\bigl(T_{\rm f}\bigr)_{\mu\nu}$ does not model a perfect fluid.  We note that $\bigl(T_f\bigr)^\mu_\nu = diag [-\rho, \, p_R \, , \, p_T \, , \,p_T]$.  The conservation of energy condition, $T^\mu_{\;\;\nu ;\mu}=0$ is vacuous for $\nu=0, 2 $ and $3$.  The only nontrivial condition is when $\nu=1$ representing the radial coordinate and in this case yields
\begin{equation}  \bigl(\, \rho + p_R \bigr)\Phi^\prime + p_R^{\; \prime} - \frac2r \, \bigl(p_T-p_R\bigr) = 0 , \end{equation} which indicates that the resultant force on a fluid element is zero.

Using the ideal gas law, $PV=nRT$, we may define the temperature per unit mass of the medium to be
\begin{equation} T \equiv \; \frac{\bar{p}}{\rho} \; \, = \; \frac13 + \frac{\Bigl(\; 1-\sqrt{1-\frac{2w(r)}r }\;\Bigr)^2}{w^\prime (r)}   \end{equation}
for free field solutions with $w(r)$ given by (22) and with the average pressure defined by $\bar{p}=(p_R+p_T+p_T)/3$.  This temperature per unit mass is dimensionless, but may be converted to a usable form by multiplying by $1.16\times 10^4\, $ degrees K per eV.

\subsection{Field with Sources.}

In spherical coordinates, a spherically symmetric tetrad with $r=\sqrt{(x^1)^2+(x^2)^2+(x^3)^2}$ may be expressed by
\begin{equation}h^i_{\;\; \mu} =  \left[ \begin{array}{cccc}
\; e^{\Phi(r)} & 0 & 0 & 0 \\
0 &\; e^{\Lambda(r)}\sin \theta \cos \phi \; & \; r\cos\theta\cos\phi \; & \; -r\sin\theta\sin\phi \; \\
0 & e^{\Lambda(r)}\sin\theta\sin\phi & r\cos\theta\sin\phi    & \; \;r\sin\theta\cos\phi \\
0 & e^{\Lambda(r)}\cos\theta\qquad & -r\sin\theta\qquad & 0
\end{array}
\right]
\end{equation}
where the upper index refers to the row. The curvature vector for this tetrad field is given by
\begin{equation} C_\mu = \frac{e^{\Lambda}}r \biggl[ \; 0,  \; 2 - e^{-\Lambda}\bigl(r\Phi^\prime + 2\bigr) , \; 0 , \;0  \biggr]  \end{equation} where components are in the order $[t,r,\theta,\phi]$ and the prime denotes the derivative with respect to $r$. The tetrad (31) leads to the metric
\begin{equation}ds^2= -e^{2\Phi(r)} dt^2 +  e^{2\Lambda(r)}dr^2+r^2d\theta^2 +r^2\sin^2\theta d\phi^2 \quad . \end{equation}
Comparison of metrics (17) and (33) implies that for the metric of (17), $(r\Phi^\prime + 2)=2e^{\Lambda}$ which then implies that $C_\mu$ in equation (32) would be identically zero.  From (32) we see that the general spherically symmetric tetrad field does not generally yield $C_\mu =0$, hence we consider whether there exists a spherically symmetric solution of the field equations which flow from (10).    The metric (33) leads to a diagonal Einstein tensor with nonzero elements:
\begin{equation}G^t_{\, t}= \frac1{r^2}\bigl(-2re^{-2\Lambda}\Lambda^\prime + e^{-2\Lambda} - 1 \bigr) = -\frac2{r^2}\frac{d}{dr}\bigl[\frac12 r (1  - e^{-2\Lambda}) \bigr] \quad , \end{equation}
\begin{equation}G^r_{\, r} = \frac1{r^2}\bigl( 2re^{-2\Lambda}\Phi^\prime + e^{-2\Lambda}-1\bigl) \end{equation}
and
\begin{equation} G^\theta_{\, \theta}=G^\phi_{\, \phi}= \frac{e^{-2\Lambda}}{r}\bigl(r\Phi^{\prime\prime}+r(\Phi^\prime)^2 - r\Phi^\prime \Lambda^\prime + \Phi^\prime - \Lambda^\prime \bigr) \quad . \end{equation}

Using $G_{\mu\nu}=8\pi T_{\mu\nu}$, we now decompose the stress-energy tensor using (13).
From $8\pi\bigl(T_{\rm f}\bigr)_{\mu\nu}= \Upsilon^{\;\;\alpha}_{\mu \;\; \nu ;\alpha}+\Upsilon^\alpha_{\;\;\sigma
\nu} \Upsilon^\sigma_{\;\; \mu \alpha} + \frac12 g_{\mu\nu}\Upsilon^{\alpha \beta
\sigma} \Upsilon_{\alpha \sigma \beta} $, one finds that $\mathbf{T}_{\rm f}$ is diagonal with elements
\begin{equation}8\pi\bigl(T_{\rm f}\bigr)_{tt}=  \frac{e^{2\Phi-2\Lambda}\bigl(   r^2\Phi^{\prime\prime}+
\frac12(r\Phi^\prime)^2 - r^2\Phi^\prime \Lambda^\prime +2r\Phi^\prime +2e^{\Lambda} -e^{2\Lambda}  -1 \bigr)}{r^2} , \end{equation}
\begin{equation}8\pi\bigl(T_{\rm f}\bigr)_{rr}=\frac{1}{r^2}\biggl(-\frac12 (r\Phi^\prime)^2 +   e^{2\Lambda}-1 \biggr)  \qquad \qquad \quad {\rm and} \qquad \end{equation}
\begin{equation}\frac{8\pi \bigl(T_{\rm f}\bigr)_{\theta\theta}}{r}= \frac{8\pi \bigl(T_{\rm f}\bigr)_{\phi\phi}}{r\sin^2 \theta} =
e^{-2\Lambda}\bigl( \frac12r(\Phi^\prime)^2 - \Phi^\prime +\Lambda^\prime +e^{\Lambda} \Phi^\prime \bigr)  \; .  \end{equation}
As indicated by (12) and (13), $\mathbf{T}_{\rm s}$ is determined by variation of the $L_s$ term in the Lagrangian (10).

\par \phantom{D} \par

\section{   Models for the Interior of a Star. }

We will use the general spherical tetrad and the field equations which are derived from the Lagrangian (10) with $L_s=\rho_s(r)$, where $\rho_s(r)$ is the density as a function of $r$.  It is well known that this Lagrangian with appropriate thermodynamic conditions lead to the usual perfect fluid stress-energy tensor [13,14].  With a tetrad that corresponds to a stationary basis (velocity of the observer is zero if $h^0_{\; \mu} = 0$ for $\mu = 1, 2 \; \mathtt{ and } \; 3$ ), one finds [12]
\begin{equation} \bigl(T_{\rm s}\bigr)^\mu_{\; \nu} = \left[
\begin{array}{cccc}
 -\rho_s \; & \; 0\; & \;0\; &\; 0\; \\
 0 & p_s & 0 & 0 \\
 0 & 0 & p_s & 0 \\
 0 & 0 & 0 & p_s
 \end{array}
 \right] \quad . \end{equation}
Using the tetrad field of (31), we require that the radial and tangential pressures of the corresponding source stress-energy tensor (11) be equal, leading to the following differential equation with primes denoting derivatives with respect to $r$:
\begin{equation} r^2\Phi^{\prime\prime} -\bigr(r^2\Lambda^\prime + r e^{\Lambda} \bigr)\Phi^\prime = 2 - 2 e^{2\Lambda} + 2r\Lambda^\prime  \end{equation}
After multiplying by an integrating factor and integrating, (41) implies that
\begin{equation}   \bigl(r\Phi^\prime + 2 \bigr)e^{-\Lambda} = 2 - \kappa re^{\int (r^{-1}e^{\Lambda}) }   \end{equation}
where $\kappa$ is arbitrary.

For convenience of interpretation we replace $\rho_s(r)$ with $\rho_s(r)-\frac1{8\pi}C^\mu_{\, ;\mu}$ and thus the new source Lagrangian term is $L_s=\rho_s(r) - \frac1{8\pi}C^\mu_{\, ;\mu}  $ . Since addition of a pure covariant divergence leaves the field equations unchanged,  this does not affect any of our conclusions thus far.  In order to leave the energy unchanged this induces the definition $p_s \equiv p+\frac1{8\pi}C^\mu_{\, ;\mu}$.  (The enthalpy [14] given by $\frac{\rho+p}{n}$, where $n$ is the baryon number density, is unchanged.)  Alternatively we may argue that we replace $C^\mu C_\mu$ by $C^\mu C_\mu + 2 C^\mu_{\; ;\mu}$ and at the same time replace $\rho_s$ with $\rho_s-\frac1{8\pi}C^\mu C_\mu$ (these changes do not affect field equations).
We also note that we assume that $C_\mu$ has compact support and is a smooth function and hence integration of the $C^\mu_{\; ;\mu}$ term over the region of support results in a value of zero and hence does not affect the overall mass as well.

With these definitions from (34-38), (41) and (42) we find that
\begin{equation}  8 \pi \rho_s   =  \frac12 \Bigl( \kappa e^{\int (r^{-1}e^{\Lambda}) } \Bigr)^2  \end{equation}
and
\begin{equation}  8 \pi p_s  =  \frac{\kappa e^{\int (r^{-1}e^{\Lambda}) } }{r} - \frac12 \Bigl( \kappa e^{\int (r^{-1}e^{\Lambda}) } \Bigr)^2    \quad .  \end{equation}
We also note that for this internal solution that the curvature vector in the order $t,r,\theta,\phi$ is given by \begin{equation} C_\mu = \Bigl[\, 0 ,\, \kappa e^{\Lambda} e^{\int (r^{-1}e^{\Lambda}) } , \; 0 , \; 0 \Bigr] \quad . \end{equation}
and this gives $C^\mu C_\mu = \kappa^2 e^{2\int (r^{-1}e^{\Lambda}) } \,$.  When the field equations are satisfied, we see that $C^\mu C_\mu = 2\rho_s = \kappa^2 e^{2\int (r^{-1}e^{\Lambda}) } $.  We conclude that the value of $C^\mu C_\mu$ is related to the density or mass of a source.


For the total stress-energy tensor $T^\mu_{\;\;\nu}$ with nonzero components given by (34-36), one finds indeed that $T^\mu_{\;\;\nu ; \mu}=0$.   From (34) with $G^t_{\; t} = - 8\pi \rho$, we also interpret the mass as a function of $r$ to be given by
\begin{equation}  - G^t_{\; t} = \frac2{r^2}\biggl(\; \frac12 r(1-e^{-2\Lambda})\;\biggr)^\prime \equiv \frac2{r^2} \biggl(\;  \frac12 m(r) \; \biggr)^\prime \end{equation}
and hence the mass within a sphere of radius $r$ is given by the function
\begin{equation}  m(r)=  r(1-e^{-2\Lambda}) \qquad . \end{equation}
This implies that $e^{2\Lambda} = (1-m/r)^{-1}$ which matches with external solution at the surface denoted by $r=R_0$.
From (35) and (47) with $G^r_{\; r}=8\pi p_R$, we get
\begin{equation} 8\pi p_R = \frac2{r^2}\biggl[ 2\sqrt{1-m/r}+\frac{3m}{2r} - 2 - \kappa r\sqrt{1-m/r}\; e^{\int\frac1{r\sqrt{1-m/r}}}     \biggr]    \end{equation}
and from (36) and (47) with $G^\theta_{\;\theta}=G^\phi_{\;\phi}= 8\pi p_T$, we get
\begin{eqnarray} 8\pi p_T & =  \frac1{r^2}\biggl[ \; \kappa\, r\, \Bigl(3\sqrt{1-m/r} - 5 \Bigr)\, e^{\int\frac1{r\sqrt{1-m/r}}}\; + \; \kappa^2 r^2 e^{\int\frac2{r\sqrt{1-m/r}}} \nonumber \\ &  \qquad \qquad \qquad \qquad \qquad \quad + \; 4(1-\sqrt{1-m/r})^2+ \frac12 m^\prime - \frac{m}{2r} \biggr]   \end{eqnarray}
There are 2 constants that may be chosen for convenience of interpretation.  The value of $\kappa$ may be determined by conditions on the pressure.  A second constant is the constant of integration in solving for $\Phi(r)$ from (42), which may be determined by appropriate continuity conditions.

{\bf Constant Density Model.}   As a reasonable model, suppose that $G^t_{\; t} = -3\alpha^2$, where $\alpha$ is an arbitrary constant and the factors of 3 is chosen for convenience.  From (46-47) we see that $m(r)= \alpha^2 r^3$  and $e^{-2\Lambda} = 1-\alpha^2 r^2$ and this model only makes sense for $0\leq r \leq 1/\alpha$.  We note that the integral that appears in (43-45) and (48-49) may be easily integrated.  Let $\hat{\kappa}= \frac{\kappa}{\alpha}$, then $\kappa \, r \, e^{\int \frac1{r\sqrt{1-m/r}}}= \hat{\kappa}\Bigl(1-\sqrt{1-\alpha^2 r^2} \, \Bigr)$ and hence $C^\mu C_\mu = \hat{\kappa}^2 \Bigl(\frac{1-\sqrt{1-\alpha^2 r^2}}r \, \Bigr)^2$.  We note that this also implies that $r\Phi^\prime = (2-\hat{\kappa})\Bigl( \, (1-\alpha^2 r^2)^{-\frac12} - 1 \Bigr)$. Integrating, we find that $\Phi(r)= \frac{C}{1+\sqrt{1-\alpha^2 r^2}}$, where constant $C$ may be chosen so that $g_{tt}$ is continuous at the surface.

In this constant density model, the resulting radial pressure is given by $8\pi p_R = \frac1{r^2}\biggl[\; 2(2-\hat{\kappa})\Bigl( \; ( 1-\alpha^2 r^2)^{\frac12}-1 \; \Bigr)\; + \;(3-2\hat{\kappa}) \alpha^2 r^2 \biggr]$.  We note that $\lim_{r\to 0}(8\pi p_R) = (1-\hat{\kappa})\alpha^2$ which suggests that a reasonable value of $\hat{\kappa}$ is less than $1$.  The radial pressure approaches a value: $8\pi p_R(r=1/\alpha)= -2(2-\hat{\kappa})\alpha^2$ which is less than zero.  At some intermediate value, it will match with the corresponding external radial pressure.   This determines the surface value, $R_0$.  If we use a result that is given in the next section, we may estimate the radial pressure at the surface to be approximately $\frac12 \alpha^2$.   Using this approximate value, we find that the radial pressure matches the external radial pressure at $r=\frac1{\alpha} \sqrt{1-\frac9{(5-4\hat{\kappa})^2}}$ and we also find that this implies that $\hat{\kappa}<1/2$.

In order to work out the value of the tangential pressure we use (49) which yields $8\pi p_T = \frac1{r^2}\biggl[ \; 2(2-\hat{\kappa})^2 \Bigl( 1 - \sqrt{1-\alpha^2 r^2} \Bigr) - (3 - 3\hat{\kappa}+\hat{\kappa}^2)\alpha^2 r^2 \biggr] $.  As $r\to 0$, $8\pi p_T \to \Bigl(1-\hat{\kappa}\Bigr) \alpha^2$ which is the same as the radial pressure.  For $r>0$, however, we see that $p_R \neq p_T$. As $r\to \frac1\alpha$, $8\pi p_T \to \Bigl( \; \hat{\kappa}^2-5\hat{\kappa}+5 \; \Bigr) \, \alpha^2$ which is positive when $\hat{\kappa}<\frac12$. Graphs of $8\pi p_R$ and $8\pi p_T$ for $\hat{\kappa}=\frac1{10}$ and $\alpha=\frac1{100}$ are shown in Figure 1.

\begin{figure}
\includegraphics[width=1.0\textwidth]{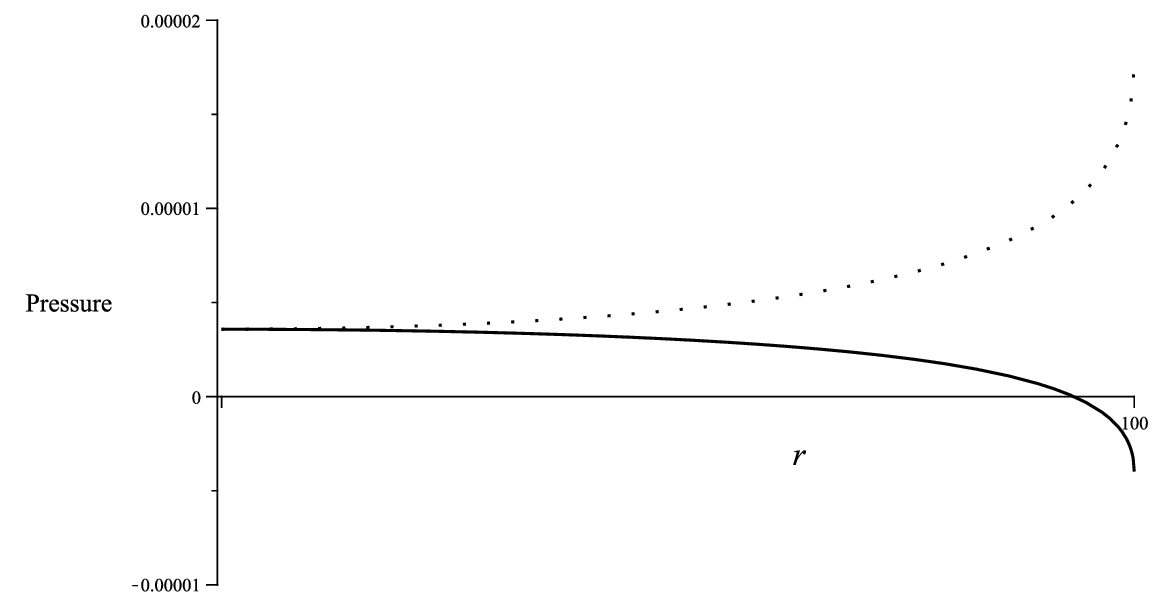}
\caption{Constant Density Model with $\hat{\kappa}=\frac1{10} \;$, $\alpha=\frac1{100}$: \quad $p_R$ (solid), $p_T$ (dotted) }
\label{fig:1}
\end{figure}

\par \phantom{D} \par \phantom{D}

\section{External Solutions.}

In order for the external solution to agree with the weak-field solution as $r\to \infty$, we will require that $\lim_{r\to\infty} w(r)=\frac12M$, where $M$ is the mass of the star as measured for very large values of $r$.   Furthermore we assume that $w(r)$ is a non-decreasing, differentiable function of $r$.  Finally, we assume that the gravitational field can be measured at the surface of the star, $r=R_0$ and hence the value of $m(R_0)$ is determined.  In general, $M \geq m(R_0)$.   These three conditions and the values $R_0$, $m(R_0)$ and $M$ will determine the boundary conditions for $w(r)$.

\subsection{An External Solution with Vanishing Density, but Non-vanishing Pressures. }
For the first example, we assume that $m(R_0)=M$ and hence choose $w(r)=\frac12M$ (in this case, this is the only admissible function for $w(r)$).  This solution also applies to the case where $m(r)$ obtains the value $M$ at a finite value, $R_1$, and then for all $r>R_1$, we have $w(r)=\frac12M$.
From (22) we have $\frac12M=\frac{r}2 -\frac{r}{2(1+\frac12r\Phi^\prime)^2}$ and hence
\begin{equation}\Phi(r) = \int \Biggl[\; \frac{2}{r\sqrt{1-\frac{M}r}}-\frac{2}{r} \, \Biggr] \; dr \end{equation}
which can be easily integrated to find $\Phi(r)=4\ln(1+\sqrt{1-\frac{M}r})+ \frac12\ln C$ for some arbitrary $C>0$.  Thus
\begin{equation} g_{tt}=-e^{2\Phi(r)}=- C \biggl(1+\sqrt{1-\frac{M}r} \biggr)^8 \quad . \end{equation}
The arbitrary constant $C$ is determined by the usual weak field approximation [12] which is $g_{tt}\approx -1 + \frac{2M}{r}$.  This implies that $C=\frac{1}{256}$.  Hence
\begin{equation}  g_{tt}=-\frac1{256}\biggl(1+\sqrt{1-\frac{M}r} \biggr)^8 \quad .  \end{equation}
We thus obtain the following line element:
\begin{equation}ds^2 = -\frac1{256}\biggl(1+\sqrt{1-\frac{M}r}\biggr)^8 dt^2 + \Bigl(1-\frac{M}r\Bigr)^{-1} dr^2 + r^2 d\theta^2 + r^2\sin^2\theta d\phi^2 \; . \end{equation}
Expanding $g_{tt}$ and $g_{rr}$ in powers of $\frac{M}r$, we find that asymptotically (for $r>>M$), to second order,
\begin{equation}ds^2 \approx -\Bigl(1-\frac{2M}{r} +\frac{5M^2}{4r^2}\Bigr)dt^2 + \Bigl(1+\frac{M}r+\frac{M^2}{r^2}\Bigr)dr^2 + r^2 d\theta^2 + r^2\sin^2\theta d\phi^2 \; . \end{equation} Using (18-20), the Einstein field equations for the external solution are
\begin{eqnarray} G_{tt} = \; 8\pi T_{tt} &= \;\; 0  \nonumber \\
G_{rr} = 8\pi T_{rr} &=\;  \frac{M \Bigl(3\sqrt{1-\frac{M}r}-1 \Bigr)}{r^3\Bigl(1-\frac{M}r \Bigr)\Bigl(1+\sqrt{1-\frac{M}r}\Bigr)} \\
\frac{G_{\theta\theta}}{r^2} = \frac{G_{\phi\phi}}{r^2\sin^2\theta} & = \frac{8\pi T_{\theta\theta}}{r^2}= \frac{8\pi T_{\phi\phi}}{r^2\sin^2\phi}  = \; \frac{-M\Bigl(9\sqrt{1-\frac{M}r}-7\Bigr)}{2r^2\Bigl(1+\sqrt{1-\frac{M}r}\Bigr)}  \quad .
\nonumber  \end{eqnarray}
Or
\begin{eqnarray}  8\pi \rho \; &= \; 0 \qquad \qquad \qquad  \nonumber \\
8\pi p_R &= \; \frac{M\Bigl(3\sqrt{1-\frac{M}r}-1 \Bigr)}{r^3\Bigl(1+\sqrt{1-\frac{M}r}\Bigr)}  \\
8\pi p_T &= \; \frac{-M\Bigl(9\sqrt{1-\frac{M}r}-7\Bigr)}{2r^2\Bigl(1+\sqrt{1-\frac{M}r}\Bigr)} \quad .
\nonumber \end{eqnarray}
Asymptotically for $r>>M$, we have \begin{eqnarray} 8 \pi p_R & \approx \;\; \frac{M}{r^3} \Bigl(1-\frac{M}{2r}\Bigr) \qquad \qquad \nonumber \\
8\pi p_T &\approx \; -\frac{M}{2r^3} \Bigl(1-\frac{2M}r\Bigr)
 \qquad. \end{eqnarray}

We note that this halo corresponds to a stressed medium since the pressures are nonzero.  Using (30), we see that the temperature per unit mass of this halo is undefined.

How do we interpret these equations?  The field equations for the space surrounding a mass of $M$ have noncompact stress-energy which is zero if and only if $M=0$, and the metric is the Lorentz metric if and only if $M=0$ as well.  The energy in this halo is a direct  consequence of the mass $M$.  This likely corresponds to dark matter or dark energy.  It is a consequence of the fact that the fundamental group of transformations is the group of conservative transformations and it has the appearance and properties associated with an actual mass or stressed medium.  Its gravitational field and effects are equivalent to that of regular matter, but it is dark in the sense that its non-gravitational effects are feeble. Its electro-weak interactions are not as dominant as the effects it has on other massive objects. Independent of the nature of the mass, we are forced to have a halo whose stress energy tensor depends on the value of $M$.

Although the stress-energy tensor for the halo does not correspond to a perfect fluid, the pressure gradients prevent the halo from moving inward or outward.  Using $T^\mu_{\;\;\nu ; \mu}=0$, with $T^\mu_{\;\;\nu} = {diag}\bigl(-\rho, p_R(r),p_T(r),p_T(r) \,  \bigr)$ one easily finds from (29) that
\begin{equation} -\frac{dp_R}{dr} + \frac2r\biggl(p_T-p_R\biggr) = \Bigl( \rho + p_R \Bigr) \Phi^\prime \quad . \end{equation}
The left-hand side of this equation corresponds to the outward force due to the pressure of the halo.  On the right-hand side, the coefficient of $\Phi^\prime$ corresponds the the inertial mass [12].  We see from (58) that the outward force due to pressures equals the inward force due to the gravitational force.  From (50), we see that asymptotically $\Phi^\prime \approx \frac{M}{r^2}$ and thus asymptotically
\begin{equation} -\frac{dp_R}{dr} + \frac2r \biggl(p_T(r)-p_R(r) \biggr) \approx  \frac{M^2}{8\pi r^5}  \quad .  \end{equation}

\subsection{An Algebraically Simple External Solution with Non-vanishing Density and Pressures.}

If the density outside (for $r\geq R_0$) is nonzero, as already noted,  ${ \lim_{r\to\infty} w(r) = \frac12 M}$.   One particularly simple model is given by \begin{equation}w(r)=\frac{M}2 - \frac{M^2}{8r} \quad . \end{equation}
With this choice, $1-\frac{2w(r)}r = 1 - \frac{M}r + \frac{M^2}{4r^2} = (1- \frac{M}{2r})^2$.  Thus, using (17), (21), (26) and (28) we have (the approximation assumes $r>>M$)
\begin{eqnarray}ds^2 &= -\Bigl(1-\frac{M}{2r}\Bigr)^4 dt^2 + \Bigl(1-\frac{M}{2r}\Bigr)^{-2} dr^2  + r^2 d\theta^2 + r^2\sin^2\theta d\phi^2  \\
&\approx -\Bigl(1-\frac{2M}r+\frac{3M^2}{r^2}\Bigr)dt^2 +\Bigl(1+\frac{M}r+\frac{3M^2}{4r^2}\Bigr)dr^2 + r^2 d\theta^2 + r^2\sin^2\theta d\phi^2 \nonumber \end{eqnarray}
and hence \begin{eqnarray} 8\pi\rho &= \frac{M^2}{4r^4}  \nonumber \\
8\pi p_r &= \frac{M}{r^3}\Bigl(1-\frac{3M}{4r}\Bigr) \\
8\pi p_T &= -\frac{M}{2r^3}\Bigl( 1 - \frac{5M}{2r}\Bigr) \nonumber \end{eqnarray}  These equations are exact.  Equation (59) is also correct for this noncompact solution.  The comments about dark matter which immediately follow equation (57) apply here as well. From (62) we see that $8\pi \bar{p}=\frac{7M^2}{12r^4}$ and hence the temperature per unit mass, determined from (30) is constant, i.e., $T=\frac73$.  Thus the halo is in thermal equilibrium.  This simple model appears to be in reasonable agreement with solar system values.  Let $0\leq \alpha<1$ represent fraction of $M$ that is due to the mass of the halo.  From (60) we see that $\alpha = \frac{M}{8R_0} \approx 2.65\times 10^{-7}$. Thus, if this model is used for the solar system, the dark matter contribution to $M$ is very small.

\subsection{Conjecture on Probability Amplitude for the Quantum Family of Manifolds.}

We claim that it is reasonable that the classical solution should correspond to a halo that is in thermal equilibrium.  We also note that for a solution that has a particular value of the Einstein tensor, that $G^\mu_{\; \nu}$ has eigenvalues that are independent of the coordinate system (if a diffeomorphism is used, then the matrix for $G^{\hat{\mu}}_{\;\hat{\nu}}$ is similar
to the matrix for $G^\mu_{\; \nu}$).  Thus the temperature per unit mass is invariant under diffeomorphisms (for example, $8\pi\rho$ is the eigenvalue associated with the only time-like eigenvector). Since we have this invariance under the diffeomorphisms, it seems reasonable to use the curvature tensor or its contractions to form the probability amplitude.  Let $R_c$ be the value of the scalar curvature for the halo which is in thermal equilibrium.  Define $\Delta R\equiv R-R_c$ for a member of the quantum family of manifolds (with $C_\mu=0$).    We conjecture that the probability amplitude that distinguishes the classical solution is $\; e^{\; i\int \Delta R \sqrt{-g}\, d^4x}$.

\subsection{Families of External Solutions for Arbitrary Values of $R_0$, $m(R_0)$ and $M$.}

In this example we exhibit a couple of families of solutions that model a gravitational field with radius of star, $R_0$, mass inside the star, $m(R_0)$, and asymptotic mass, $M$.  As above we will let $\alpha$ represent the fraction of $M$ that is due to the mass of the halo (or dark matter) and hence $m(R_0)=(1-\alpha)M$.

{\it Linear Model.}  Let $k=\frac{(1-\alpha)M}{R_0}$.  Using (30), we note that the linear function $w(r)=\frac{(1-\alpha)M}{2R_0} \, = \, \frac12 k r$ results in a constant temperature per unit mass of $T=\frac13 + \frac{2\Bigl( \; 1- \sqrt{1-k} \;\Bigr)^2}{k} \approx \frac13 + \frac{k}2$ when $R_0 >> M$.
Thus we define
\begin{equation}
w(r)= \Biggl\{
\begin{array}{cc}
 \frac{(1-\alpha)M}{2R_0} \, r \quad &  ,  R_0\leq r \leq \frac{R_0}{1-\alpha} \\
\frac{M}2  \quad & , \; r > \frac{R_0}{1-\alpha} \qquad
 \end{array}
\end{equation}
According to this model, the halo extends to $r=\frac{R_0}{1-\alpha}$.  For typical value of $\alpha$ such as $0.05$, this yield a halo radius of $20R_0$.  This may be a model that could be used for halos of galaxies since the radial velocity curve for particles in circular orbits would be constant.  For this model the line element is (for $R_0\leq r \leq \frac{R_0}{1-\alpha}$)
\begin{equation} ds^2 = -Cr^{\frac4{\sqrt{1-k}}-4} dt^2 + \Bigl( 1-k \Bigr)^{-1} dr^2  + r^2 (d\theta^2 + \sin^2\theta d\phi^2)  \end{equation}
where $C$ is a constant chosen so that the metric will match the metric of (53) when $r=\frac{R_0}{1-\alpha}$.  For $r>\frac{R_0}{1-\alpha}$, all quantities match that of (53-56).
The density and pressures are (for $R_0\leq r \leq \frac{R_0}{1-\alpha}$)

\begin{eqnarray} 8\pi\rho &=  \frac{k}{ r^2}            \nonumber \\
8\pi p_r &= \frac1{r^2}\Bigl(1-\sqrt{1-k}\Bigr)\Bigl(3\sqrt{1-k}-1\Bigr) \\
8\pi p_T &= \frac4{r^2}\Bigl(1-\sqrt{1-k}\Bigr)^2 \nonumber \end{eqnarray}

{\it Rational Function Model.}  Another choice is the rational function \begin{equation} w(r)=\frac{M}2\Bigl( \; 1-\frac{\alpha R_0}r \;\Bigr) \;  \end{equation}
(recall $\; m(R_0)=(1-\alpha)M\;$).   For large $r$,  $\, T\approx \frac13 + \frac{M}{2\alpha R_0}$.  Thus the temperature per unit mass is approximately constant.  The halo extends indefinitely in this model and we also find that the density is given by $8\pi \rho = \frac{\alpha M R_0}{r^4}$.  The line element, density and pressures are easily calculated from (21-28).

\par \phantom{D} \par

\section{ Motion of a Test Particle in the External Field Solution. Comparison with Solar System Predictions of General Relativity. }

We now investigate the motion of a test particle in the external field solution.  We will develop general formulas and primarily apply them to the metrics (53) and (61). These metrics seem to be the ones that a suitable for solar system applications (dark matter is not a significant portion of the total mass).  We assume that $\alpha$, the fraction of total mass $M$ due to the halo, is small so that if the correct model is the multipart linear model of (63), then almost all of the motion that we are analyzing is beyond $r=\frac{R_0}{1-\alpha}$ where (53) applies.

We emphasize that the results of this section, while consistent with the results of this paper, are tentative and likely a rough estimate to a rigorous application of our theory.  Arguments are given that show that our theory could be the correct theory even though at first inspection it would appear otherwise.  The most important issue affecting the application of our theory to test particles is the fact that motion takes place within a stressed medium with a nonzero stress-energy tensor.

An efficient procedure for finding equations of motion is one that extremalizes an appropriate Lagrangian.  We follow de Felice and Clarke [14] with a Lagrangian for a particle in a field with nonzero stress energy tensor.  Specifically,  we use the Lagrangian (10) with the source term given by
\begin{equation}  L_s = \rho(x)  = \mu \int \delta_\epsilon^4(x-\gamma(s))(-u^\mu u_\mu)^{\frac12} ds   \end{equation}
where $\delta_\epsilon^4$ approximates the Dirac delta function with a space-like volume of $\epsilon$ which yields the usual Dirac delta function in the limit as $\epsilon\to 0$.  The path of the particle is given by $\gamma(s)$ and its velocity is $u^\mu=\frac{dx^\mu}{d\tau}$.  For convenience, we will use the "dot" notation for the components of $u^\alpha$, i.e. $u^\alpha = \langle \dot{t},\dot{r},\dot{\theta},\dot{\phi}\rangle$.  Let $\mu$ denote the mass of the particle. As noted in de Felice and Clarke ([14] page 222), the condition  $T^{\beta \alpha}_{\; \;\;\; ; \beta }=0$ leads to
\begin{equation}  \epsilon \, (T_f)^{\beta\alpha}_{\;\;\;\; ;\beta} \; + \; \frac{\mu}{\sqrt{-u^\nu u_\nu}} \, u^\beta u^\alpha_{\; ;\beta} = 0  \qquad . \end{equation}
The only nonzero component of $(T_f)^{\beta\alpha}_{\;\;\;\; ;\beta}$ is the radial component ($\alpha=1$) as we saw in (29).  The $u^\beta u^\alpha_{\; ;\beta}$ term corresponds to the geodesic equation.  When $\alpha \neq 1$, (68) is equivalent to the geodesic equation for $u^\alpha$.

First we note that the $\theta$ component (when $\alpha=2$) of (68) after multiplying by $ \frac{\sqrt{-u^\nu u_\nu}}{\mu} $  yields
\begin{equation}   \frac{d^2 \theta}{d \tau^2} + \frac2r \frac{dr}{d\tau}\frac{d\theta}{d\tau} - (\sin\theta \cos\theta)\biggl(\frac{d\phi}{d\tau}\biggr)^2=0  \qquad . \end{equation}  We note that $\theta\equiv \frac{\pi}2$ is a solution of this equation and symmetry considerations imply that we may safely assign this value of $\theta$ since particle motion takes place in a plane through the origin ($r=0$).

We next look at the $t$ component (when $\alpha=0$) of (68)  which yields:
\begin{equation} \frac{d^2t}{d\tau^2} + \frac{4M}{r^2\sqrt{1-M/r}\Bigl(1+\sqrt{1-M/r}\Bigr)}\frac{dr}{d\tau}\frac{dt}{d\tau} = 0   \qquad .   \end{equation}
After multiplying by an integrating factor we find that this equation may be written as $\frac{d}{d\tau}\Bigl[\frac1{256}(1+\sqrt{1-M/r})^8\frac{dt}{d\tau}\Bigr]=0$ and hence
\begin{equation}  \dot{t} = 256E\Bigl( \, 1+\sqrt{1-M/r}\;\Bigr)^{-8}  \end{equation} where $E$ is a constant representing the energy of the particle.

The $\phi$ component (when $\alpha=3$) of (68) yields:
\begin{equation}   \frac{d^2 \phi}{d\tau^2} \; + \; \frac2r \frac{dr}{d\tau} \frac{d\phi}{d\tau} = 0      \end{equation}
and after multiplying by $r^2$ this equation may be written as $\frac{d}{d\tau}\Bigl[ r^2\frac{d\phi}{d\tau}\Bigr]=0$.  Thus
\begin{equation}  \dot{\phi} =  \frac{L}{r^2}  \end{equation}
where the constant $L$ represents the angular momentum which is conserved.

The $r$ component requires some careful interpretation. We will assume the the particle is small in the sense that the curvature of space does not change appreciably over the space-like regions associated with its motion. We will also assume that the particle has spherical symmetry.  Thus there is an external field  associated with the particle that is carried along with it (halo).  It seems reasonable that the pressures in this particle halo, similar to those of equations (56) or (62), will have very little effect on the motion of the particle.   The density of the particle's halo will be incorporated into the calculation of the mass $\mu$ of the particle.  If the density of the particle is identical to the density determined by  $(T_{\rm f})^\mu_{\;\nu}$, i.e. equal to $ (T_{\rm f})^0_{\; 0}$, (i.e. identical to the density of the fluid elements of the halo associated with the mass $M$) then the net force on the particle would be zero.  However, we find that when the density differs from the fluid element density, then the $(T_{\rm f})^{\beta\alpha}_{\;\;\; ; \beta}$ term has a nonzero contribution.  As is usual for the perfect fluid type stress-energy tensor, the components of $(T_{\rm f})^{\beta\alpha}_{\;\;\; ; \beta}$ are in units of force per unit volume.   We find that the gravitational action on the particle is accounted for in the $\frac{\mu}{\sqrt{-u^\nu u_\nu}} \, u^\beta u^\alpha_{\; ;\beta}$ term of (68).  Thus, the corresponding term of $(T_{\rm f})^{\beta\alpha}_{\;\;\; ; \beta}$ should be omitted.  (Recall that $\epsilon$ represents the volume of the particle.)  This implies that there is an additional outward force, $F_p(r)$, from the radial and tangential pressures, since
\begin{eqnarray} \epsilon \, (T_f)^{\beta r}_{\;\;\;\; ;\beta} \; + \; \frac{\mu}{\sqrt{-u^\nu u_\nu}} \, u^\beta u^r_{\; ;\beta} \; & =  \; 0   \nonumber \\
\epsilon \, \biggl( \, p_R^{\; \prime} - \frac2r \, \bigl(p_T-p_R\bigr)\,\biggr) \; + \; \frac{\mu}{\sqrt{-u^\nu u_\nu}} \, u^\beta u^r_{\; ;\beta}  & = \; 0  \end{eqnarray}
i.e. \begin{equation} \frac{\mu}{\sqrt{-u^\nu u_\nu}} \, u^\beta u^r_{\; ;\beta} \; = \; \epsilon \, \biggl( -p_R^{\; \prime} + \frac2r \, \bigl(p_T-p_R\bigr)\, \biggr) \; \; \equiv \; F_p   \quad . \end{equation}

The $r$ component of (68) (when $\alpha=1$)after multiplying by $  \frac{\sqrt{-u^\nu u_\nu}}{\mu}  $ is given by
\begin{eqnarray} \ddot{r} + \frac{M\Bigl(1+\sqrt{1-M/r}\Bigr)^7\sqrt{1-M/r}}{128r^2} \, \dot{t}^2 & \nonumber \\ \quad \; - \; \frac{M}{2r^2(1-M/r)}\, \dot{r}^2   - r(1-M/r)\dot{\phi}^2 \; & = \;  \frac{\sqrt{-u^\nu u_\nu}}{\mu} \, F_p \end{eqnarray}
Using (71) and (73) we find that
\begin{eqnarray} \ddot{r} + \frac{512 M\,E^2\sqrt{1-M/r}}{r^2\Bigl(1+\sqrt{1-M/r}\Bigr)^9}\, - \; \frac{M}{2r^2(1-M/r)} \, \dot{r}^2  & \nonumber \\ \qquad \qquad \qquad \qquad \qquad \qquad \;\; - \;\; \frac{L^2(1-M/r)}{r^3}  \; & = \;  \frac{\sqrt{-u^\nu u_\nu}}{\mu} \, F_p \end{eqnarray}
We now impose a normalization on the velocity $u^\mu$: \quad $-u^\nu u_\nu \approx 1$.  (It actually should be $-u^\nu u_\nu - g_{rr}\int \frac{F_p}{\mu} = 1$ with the constant of integration chosen so that the  integral term vanishes as $r\to \infty$.  This correction to $-u^\nu u_\nu$ is much less than $\frac{M}r$. Furthermore it is multiplied by $\frac{F_p}{\mu}$ which is also small.)  Therefore $\frac1{256}\Bigl(1+\sqrt{1-\frac{M}r}\Bigr)^8 \dot{t}^2-\bigl(1-\frac{M}r\bigr)^{-1}\dot{r}^2 -r^2\dot{\phi}^2 =1$.  Using (71) and (73) we may eliminate the $\dot{t}$ and $\dot{\phi}$ terms.  This leads to $256E^2\Bigl(1 +\sqrt{1-\frac{M}r}\Bigr)^{-8}=\bigl(1-\frac{M}r\bigr)^{-1}\dot{r}^2+\frac{L^2}{r^2}+1$.  Substituting this into (77), we arrive at
\begin{eqnarray} \ddot{r} \; + \; \frac{2M\sqrt{1-\frac{M}r}}{r^2\Bigl(1+\sqrt{1-\frac{M}r}\Bigr)} \; & - \; \frac{L^2\sqrt{1-\frac{M}r}\Bigl(3\sqrt{1-\frac{M}r}-2\Bigr)}{r^3} \nonumber \\ & + \; \frac{M\Bigl(3\sqrt{1-\frac{M}r}-1\Bigr)}{2r^2\Bigl(1-\frac{M}r\Bigr)\Bigl(1+\sqrt{1-\frac{M}r}\Bigr)}\dot{r}^2  \;\; = \;  \frac1{\mu} F_p \;\;\;   .   \end{eqnarray}
A similar computation with the metric given by (61) results in
\begin{equation}  \ddot{r} \; + \; \frac{M(1-\frac{M}{2r})}{r^2} + \frac{M}{2r^2(1-\frac{M}{2r})} \dot{r}^2 - \frac{L^2(1-\frac{M}{2r})(1-\frac{3M}{2r})}{r^3} \; = \; \frac1{\mu} \, F_p   \quad .    \end{equation}
From (59) we see that $\frac1{\mu} \, F_p  \approx \frac{\epsilon}{\mu} \frac{M^2}{8\pi r^5}$.  Let the average density of the particle be given by $\tilde{\rho}$, then $\tilde{\rho}=\frac{\mu}{\epsilon}$ and so we see that
\begin{equation}  \frac1{\mu} \, F_p \approx   \frac{M^2}{8\pi \tilde{\rho} r^5} \qquad . \end{equation}
The mean radius of the earth is $6.3675 \times 10^8$ cm with a mass in geometrized units of 0.4438 cm.  This yield a value of $\tilde{\rho}$ of approximately $4.0971 \times 10^{-28} {\rm cm}^{-2}$.  Typical values of $\tilde{\rho}$ for planets range between $3 \times 10^{-29} {\rm cm}^{-2}$  and $5\times 10^{-28} {\rm cm}^{-2}$.  Consider the ratio of $F_p$ to the magnitude of the (Newtonian) gravitational force of the sun, $F_{\rm grav}\equiv \frac{\mu M}{r^2}$.  This ratio is $\frac{F_p}{F_{\rm grav}} \approx \frac{M}{8\pi\tilde{\rho}r^3}$. For the planet Mercury this ratio is approximately $7.53\times 10^{-8}$.  Table 1 gives values of $\; \frac1{\mu}F_p$,  $\frac{F_p}{F_{\rm grav}}$, $\frac{M}{r}$ and  $\frac{M^2}{r^3}$.

\begin{table}
\caption{\label{arttype}  Values of $F_p$, $\frac{M}{r}$ and $\frac{M^2}{r^3}$ for various planets}
\footnotesize\rm
\begin{tabular}{|l|l|l|l|l|}\hline  
Planet \phantom{$4^{X^{X^X}}_{X_{X_X}}$} & \phantom{DD}$\; \frac1{\mu}\,F_p$& \phantom{DD} $\; \frac{F_p}{F_{\rm grav}}$& \phantom{DD} $\; \frac{M}{r}$ & \phantom{DD} $ \; \frac{M^2}{r^3}$ \\
\hline 
 Mercury \phantom{$4^{X^X}$}  &$3.31\times 10^{-28} {\rm cm}^{-1}$&$7.53\times 10^{-8}$&$2.55\times 10^{-8}$ & $1.12\times 10^{-28} {\rm cm}^{-1}  $  \\
Earth \phantom{$4^{X^X}$} &$2.82\times 10^{-30} {\rm cm}^{-1}$&$4.28\times 10^{-9}$&$9.86\times 10^{-9}$ & $6.51\times 10^{-30} {\rm cm}^{-1}$\\
Jupiter \phantom{$4^{X^X}$} &$3.08\times 10^{-33} {\rm cm}^{-1}$&$1.26\times 10^{-10}$&$1.90\times 10^{-9} $ & $4.62\times 10^{-32} {\rm cm}^{-1}$ \\
Neptune \phantom{$4^{X^X}$} &$3.60\times 10^{-37} {\rm cm}^{-1}$&$4.94 \times 10^{-13}$&$3.28\times 10^{-10} $ &  $2.39\times 10^{-34} {\rm cm}^{-1}$ \\
\hline
\end{tabular}
\end{table}

{\bf Kepler's Law.}  The angular velocity is given by $\omega = \frac{\;\dot{\phi}^{\phantom{T}}}{\;\dot{t}^{\phantom{T}}}\;$, and so when the orbit is circular ($\ddot{r}=\dot{r}=0$) we see generally that (75) and  the normalization $-u^\alpha u_\alpha = 1$ imply
\begin{equation} \Gamma^r_{tt} \dot{t}^{\,2} + \Gamma^r_{\phi\phi} \dot{\phi}^{\,2} = -\frac1{\mu}\, F_p \, \Bigl( \, g_{tt} \, \dot{t}^{\,2} + g_{\phi\phi} \, \dot{\phi}^{\,2} \, \Bigr) \quad .  \end{equation}
Solving for $\omega^2$ and multiplying by $r^3$ yields
\begin{equation}  r^3\omega^2 = \;\frac{ -r^3\Bigl( \, \Gamma^r_{tt} +\frac1{\mu}F_p \, g_{tt}\Bigr)}{\Gamma^r_{\phi\phi} + \frac1{\mu} F_p \, g_{\phi\phi} } \end{equation}
We will assume that $\frac{M}{r}<<1$ and that $\frac{F_p}{F_{\rm grav}} \approx  \frac{M}{8\pi\tilde{\rho} r^3}$ is small and is approximately the same size as $\frac{M}r$.  These assumptions are supported by the values in Table 1. For  the metric of (53) we find that $\Gamma^r_{tt} = \frac{M}{128r^2} \Bigl(1+\sqrt{1-M/r}\Bigr)\,\sqrt{1-M/r}$ and $\Gamma^r_{\phi\phi}= -r\Bigl(1-\frac{M}{r}\Bigr)$.  Thus using this and (53) we find
\begin{equation}\omega^2 r^3 \approx M\Bigl(  1-\frac{5M}{4r}-\frac{M}{8\pi \tilde{\rho} r^3} \Bigr)  \quad .  \end{equation}
For the motion under the metric (61) one gets $\Gamma^r_{tt}= \frac{M}{r^2}\Bigl(1-\frac{M}{2r}\Bigr)^5$ and $\Gamma^r_{\phi\phi} = - r\Bigl(1-\frac{M}{2r}\Bigr)^2  $  and hence under these assumptions we find \begin{equation}\omega^2 r^3 \approx M\biggl(1-\frac{3M}{2r} - \frac{M}{8\pi \tilde{\rho} r^3}\biggr) \quad .  \end{equation}
Thus we see that when $\frac{M}{r}$ and $\frac{F_p}{F_{\rm grav}}$ as very small that we have excellent agreement with Kepler's Law.

{\bf Radial Motion.}  For pure radial motion ($L=0$), (78) with $r>>M$ asymptotically yields  \begin{equation} \ddot{r}\approx -\frac{M}{r^2}\Bigl(1-\frac{M}{4r}\Bigr) - \frac{M}{2r^2}\Bigl(1+\frac{M}{2r}\Bigr) \dot{r}^2 + \frac{M^2}{8\pi \tilde{\rho} r^5}  \quad  ,  \quad .    \end{equation}
From the metric (61), one finds from (79) with $r>>M$, that the pure radial motion to be approximately given by
\begin{equation} \ddot{r} \approx - \frac{M}{r^2}\biggl(1-\frac{M}{2r}\biggr) - \frac{M}{2r^2}\biggl(1+\frac{M}{2r}\biggr)\dot{r}^2 + \frac{M^2}{8\pi \tilde{\rho} r^5} \quad . \end{equation}
The magnitude of the $\dot{r}^2$ terms in (85-87) do not appear to be large enough to explain the Pioneer anomaly. The Pioneer spacecraft is traveling out of the solar system.  A small acceleration toward the sun which cannot be explained by general relativity has been observed over a period of years [15].  For Pioneer, the magnitude of these terms in (85 - 86) at planet Pluto is approximately $10^{-15}$ m s$^{-2}$ which is much less that the anomalous value of about $8.74\times 10^{-10}$ m s$^{-2}$.

However, we do see that there is an explanation of the Pioneer anomaly.  These radial equations  (85-86) have additional outward accelerations that are not part of the standard external Schwarzschild solution equations.  From (85) we see an additional outward acceleration given by
\begin{equation} a_{\rm out} = \frac{M^2}{4r^3}+\frac{M^2}{8\pi\tilde{\rho} r^5} \qquad . \end{equation}
However the $\tilde{\rho}$ value is for the Pioneer spacecraft instead of the planet's mean density.  A rough estimate of the volume of the Pioneer spacecraft is $540,000\; {\rm cm}^3$ for the main compartment and approximately $200,000 \; {\rm cm}^3$ for the remaining components (note: this is a rough estimate).  Thus $\tilde{\rho} \approx \frac13 {\rm g/cm}^3$.  This yields $\frac{M^2}{8\pi \tilde{\rho} r^5}\approx 4.67 \times 10^{-29} {\rm cm}^{-1}$ at a 1 A.U. from the sun.
At Earth, we see using the value of $\frac{M^2}{r^3}$ from Table 1, that $a_{\rm out} \approx 0.25 \times 6.51\times 10^{-30} + 4.67 \times 10^{-29} \, {\rm cm}^{-1}$.  Thus the Pioneer spacecraft at Earth's distance from the sun has an outward acceleration of
\begin{equation}  a_{\rm out} \approx  4.83 \times 10^{-29} {\rm cm}^{-1} \qquad {\rm ( at\;\; Earth)}. \end{equation}  This is an extra outward acceleration due to the fact that our theory differs from general relativity and also includes a nonzero stress-energy tensor.  At a distance of Jupiter from the Sun, with the same value of $\tilde{\rho}$ yields $\frac{M^2}{8\pi \tilde{\rho} r^5} \approx 1.23 \times 10^{-32} {\rm cm}^{-1}$.  Thus, using the value of $\frac{M^2}{r^3}$ from Table 1, we see that
\begin{equation} a_{\rm out} \approx 1.39\times 10^{-32} {\rm cm}^{-1} \qquad {\rm (at \;\; Jupiter\, )} \, .  \end{equation}
For distances that are greater than the distance from the Sun to Jupiter we see that $\Delta a_{\rm out} \approx 4.83 \times 10^{-29} {\rm cm}^{-1}$ and under the general relativity model, this would be interpreted as an additional Sun-ward acceleration.  Converting this value to standard units yields
\begin{equation}  \Delta a_{\rm out} \approx 4.34 \times 10^{-10} \; {\rm m \, s^{-2} }   \end{equation}
which is about 50\% of the anomalous acceleration.
For the metric (61) at earth we have $a_{\rm out} \approx 4.99 \times 10^{-29} {\rm cm}^{-1}$ which yields a value of
\begin{equation}  \Delta a_{\rm out} \approx   4.49 \times 10^{-10} \; {\rm m \, s^{-2}}  \end{equation}
which is 51\% of the anomalous acceleration. The remaining anomalous acceleration may be explained by thermal forces [16].

{\bf Redshift.}  The difference between the values of $g_{tt}$ in this model and the standard Schwarzschild solution would produce small differences in the predicted redshift.  The redshift $z=\frac{\Delta \lambda}{\lambda} = |g_{tt}|^{-\frac12}-1$ for stationary objects.  From (51) we find that \begin{equation}z=16\biggl(1+\sqrt{1-\frac{M}r} \; \biggr)^{-4} -1 \; \; \approx  \frac{M}{r} + \frac{7M^2}{8r^2} \quad , \quad r >> M \quad ,  \end{equation} and from (61) we find
\begin{equation} z=  \biggl(1-\frac{M}{2r}\biggr)^{-2} - 1 \; \; \approx \frac{M}{r} + \frac{3M^2}{4r^2}  \quad , \quad r>>M \quad .  \end{equation} Asymptotically, these results agree with the value found in the Schwarzschild geometry, i.e. $z\approx \frac{M}r$.
At the distance of the earth from the sun, one finds the value given by (92) differs from the standard value by $8.5\times 10^{-17}$, with a relative difference of $8.6\times 10^{-9}$.  From (93) we find the value of $z$  differs from the standard value by $7.3 \times 10^{-17}$ with a relative difference of $7.4 \times 10^{-9}$.

{\bf Precession of Perihelion.}  We now consider the precession of perihelion problem.   Assuming spherical symmetry and using the $-u^\alpha u_\alpha=1$ normalization, we have $g_{tt}\dot{t}^2+g_{rr}\dot{r}^2+r^2\dot{\phi}^2=-1$ with motion restricted (without loss of generality) to the $\theta=\frac{\pi}2$ plane.  Now $\dot{t}=-g^{tt}E$ and $\dot{\phi}=\frac{L}{r^2}$.  Hence $g^{tt}E^2+g_{rr}\dot{r}^2+\frac{L^2}{r^2} = - 1$.  After differentiating this equation we see that equations (78) and (79) are not recovered unless a term is added, specifically, we get $g^{tt}E^2+g_{rr}\dot{r}^2+\frac{L^2}{r^2} - g_{rr}\int \frac{F_p}{\mu} = - 1$.  Using the approximation $\frac{F_p}{\mu}$ in (80) we find
\begin{equation}  \dot{r}^2 = -\biggl(\frac{g^{tt}}{g_{rr}}\biggr)E^2 - \frac1{g_{rr}}\Bigl(1+\frac{L^2}{r^2}\Bigr) - \frac{M^2}{16\pi \hat{\rho} r^4}  \end{equation}
Using $\frac{dr}{d\phi} = \frac{r^2}{\tilde{L}} \frac{dr}{d\tau}$, with $u\equiv \frac{M}r$ and $L^\dagger \equiv \frac{L}{M}$, one finds that \begin{equation}\Bigl(L^\dagger \frac{du}{d\phi} \Bigr)^2 \; =
\, \Bigl(\frac{-g^{tt}}{g_{rr}} \Bigr) E^2 - \frac1{g_{rr}} \Bigl( 1 + (L^\dagger)^2 u^2 \Bigr) - \frac1{16\pi\hat{\rho}M^2} u^4 \; \equiv \; f(u)
\quad .  \end{equation}
When $E$ is large $f(u)>0$ and the value of $u$ oscillates. When the orbit is circular at $u=u_0=\frac{M}{r_0}$, the function $f(u)$ has a maximum with both $f(u_0)=0$ and $f^\prime(u_0)=0$.   Hence $f(u) \approx \frac12 f^{\prime\prime}(u_0) (u-u_0)^2$.   Via the chain rule, one has $2(L^\dagger)^2\frac{du}{d\phi}\frac{d^2u}{d\phi^2} = f^\prime(u) \frac{du}{d\phi}$.  Thus, one finds that, \begin{equation}\frac{d^2}{d\phi^2}\biggl(u-u_0\biggr) - \frac{f^{\prime\prime}(u_0)}{2(L^\dagger)^2} \Bigl(u-u_0\Bigr) = 0 \quad .  \end{equation}
When $f^{\prime\prime}(u_0)<0$, the solution is periodic with \begin{equation} Period =\frac{2\pi}{\sqrt{-\frac{f^{\prime\prime}(u_0)}{2(L^\dagger)^2}}} \quad .  \end{equation}
From the metric given in (53), we find that \begin{equation}{-\frac{f^{\prime\prime}(u_0)}{2(L^\dagger)^2}}\approx 1 - \frac{15}4 u_0 + \frac{12 u_0^{\; 2}}{16\pi\hat{\rho}L^2} \qquad . \end{equation}
For Mercury, the $u_0^{\; 2}$ term (last term) of this expression is approximately $10^{-4}$ times the value of the preceding term and thus the perihelion is shifted by \begin{equation} \Delta \phi  \approx \biggl(\frac{15\pi}{4}\biggr) \frac{M}{r_0} \quad . \end{equation}
where $r_0$ is the radius of the near-circular orbit.  If the metric (61) is used one finds
\begin{equation}{-\frac{f^{\prime\prime}(u_0)}{2(L^\dagger)^2}}\approx 1 - \frac{7}2 u_0 + \frac{12 u_0^{\; 2}}{16\pi\hat{\rho}L^2} \qquad . \end{equation}
and hence
\begin{equation}\Delta \phi \approx \biggl(\frac{7\pi}{2}\biggr) \frac{M}{r_0} \quad .  \end{equation}
Both of these results are less than the standard result of $\frac{6\pi M}{r_0}$, with (99) being $
\frac58$ of the standard result and (101) being $\frac7{12}$ of the standard result.  In the Newcomb's calculation of the precession of Mercury in 1882, [17], it was stated that "a planet or a group of planets between Mercury and the Sun" could explain the additional $43.03^{\prime\prime}$ per century.  It seems reasonable to define the average pressure by $\bar{p}=\frac13\bigl(p_r+2p_T\bigr)$ and thus the inertial mass per unit volume of the halo is given by $\rho+\bar{p}$.  The inertial mass of the halo between the Sun and Mercury for the metric given by (53) is given by
\begin{equation} \Delta m \approx \int_{r_{sun}\leq r \leq r_{merc}} \frac{M^2}{8\pi\cdot2r^4} \, d^3x \approx  0.0784 \; {\rm cm} \quad . \end{equation}
This is about 17.7\% of the mass of Earth. For the metric of (61), the inertial mass is $\frac53$ times larger, giving 0.131 cm which represents 29.4\% of the Earth's mass.  It is possible that these values of $\Delta m$  may explain the remaining fraction of the anomalous precession that is not explained by (99) and (101). Since the mass halo is spherically symmetric, other effects on the orbit of Mercury should be minimal.

{\bf Isotropic Form and Temperature of the Corona.}  For most problems in astrophysics, the isotropic form of the metric is preferred.  From the metric of (15),  which is generated by tetrad of (14), the field equations $C_\mu = 0$ are satisfied.  For the weak field approximation to hold, $f(r)\approx 1-\frac{2M}{r}$ for large $r$.
The tetrad that produces the isotropic form for the metric given in (61) above, is generated by the transformation, $r\to r+\frac{M}2$.  The resulting metric is
\begin{equation} ds^2 \, = \, -\biggl(1+\frac{M}{2r}\biggr)^{-4} dt^2 + \biggl(1+\frac{M}{2r}\biggr)^2\Bigl[\; dr^2 + r^2 d\theta^2 + r^2 \sin^2\theta d\phi^2\;\Bigr] \end{equation}
with
\begin{eqnarray} 8\pi\rho &= \frac{M^2}{4r^4(1+\frac{M}{2r})^4}  \nonumber \\
 8\pi p_r &= \frac{M}{r^3(1+\frac{M}{2r})^4}\biggl(\;1-\frac{M}{4r}\;\biggr) \\
 8\pi p_T &= \frac{-M\;\;}{2r^3(1+\frac{M}{2r})^4}\biggl(\; 1 - \frac{2M}{r}\; \biggr) \nonumber \end{eqnarray}
To first order, the isotropic metric and its corresponding stress energy tensor are equivalent to the metric of (61).

We again note that, unlike most other alternatives to general relativity, the theory based on the conservation group when interpreted as a manifold has a non-vanishing stress energy tensor.  Suppose we apply (30) and assume that the halo is comprised of particles with mass $\hat{m}$ eV.  For the metric (61), the resulting temperature per unit mass is $T=\frac73$, i.e., $ 2.70\times 10^4\, $ degrees Kelvin per electron volt.   If the masses of the constituent matter in the halo are approximately 36 ev, the resulting temperature would be approximately $10^6\;$K and hence would explain the high temperature of the corona.  This suggests that dark matter is composed of particles of small mass, possibly a mixture of the neutrinos $\nu_e$, $\nu_\mu$ and $\nu_\tau$.

{\bf Deflection of Light and Time Delay. }  For null rays which model photon motion, $ds^2=0$, and we see from (15) that for position vector ${\bf r}$,
\begin{equation} \biggl|\frac{d{\bf r}}{dt}\biggr| = \bigl[f(r)\bigr]^\frac34 \approx \biggl(\; 1-\frac{2M}r\; \biggr)^\frac34  \approx \; 1-\frac{3M}{2r}\; \approx \frac1{1+\frac{3M}{2r}} \qquad . \end{equation}
The denominator of the last expression in (105) represents a refraction index of $n_0(r)\equiv 1+\frac{3M}{2r}$.  This is a general result and, as a check, one easily sees that the isotropic metric given in (103) satisfies this condition.  We see that $n_0(r)-1=\frac{3M}{2r}$ is is precisely 75\% of the value in general relativity (where $n(r)-1 = \frac{2M}{r}\,$).
In our theory, however, the resulting density and pressures (103) indicate a stressed medium through which the electromagnetic radiation passes.  Let $\bar{p}=\frac13\bigl(p_r+p_T+p_T\bigr)$ be the average pressure.  We propose that additional refraction occurs due to the medium and the value of $n_1(r)-1$ is proportional to $\rho + \bar{p}$,  viz.
\begin{equation}  n_1(r) - 1 = 8\pi\alpha_r(\rho+\bar{p}) \quad , \end{equation}
where $\alpha_r$ is a positive constant and the factor of $8\pi$ is included for convenience.  This formula may be justified by the
Lorentz-Lorenz relation [18].

For the deflection of light problem we will follow the analysis of de Felice and Clarke [14, p 354].  We see from the stress energy tensor (57) that $8\pi\bigl(\rho+\bar{p}\bigr) \approx \frac{M^2}{2r^4}$, and for the stress energy tensor of (62), $8\pi\bigl(\rho+\bar{p}\bigr) \approx \frac{5M^2}{6r^4}$.  The values computed from the isotropic form of the metric are the same to the order of approximation used.   Hence $n_1(r)\approx 1+\frac{\beta_r M^2}{r^4}$,  with $\beta_r=\frac12\alpha_r$ for (57) and $\beta_r=\frac56 \alpha_r$ for the stress energy tensor (62).  We multiply this by the corresponding refractive index which is calculated from the metric, hence
\begin{equation}  n(r) = n_0(r) \cdot n_1(r) \approx 1 + \frac{3M}{2r} + \frac{\beta_r M^2}{r^4} \quad . \end{equation}
From [14] we find that the angle of deflection of light passing near the surface of the sun (i.e. a minimum radius of $r_0$) is given by
\begin{equation} \Delta \phi =  \int_{1+\frac{3M}{2r_0}+\frac{\beta_r M^2}{r_0^4}}^\infty \, \frac{2\, dr}{r\sqrt{r^2-1}}  \; - \; \int_1^\infty \frac{2\, dr}{r\sqrt{r^2(1+\frac{3M}{2r}+\frac{\beta_r M^2}{r^4})^2-1}}  \end{equation}
Defining $A\equiv \frac{3M}{2r_0}$, $B\equiv \frac{\beta_r M^2}{r_0^4}$ and changing variables to $w\equiv \frac{r}{r_0}$ we find that \begin{equation} \Delta \phi = \int_{1+A+B}^\infty \, \frac{2\, dw}{w\sqrt{w^2-1}} \; - \; \int_1^\infty \frac{2\, dw}{w\sqrt{w^2(1+\frac{A}w+\frac{B}{w^3})^2-1}} \end{equation}
Noting that $A$ and $B$ are much less than 1, we find that (109) yields
\begin{equation} \Delta \phi \approx -2A - \frac{3\pi}2B +(2A+8B)\sqrt{2(A+B)} \end{equation}

In the calculation of the time delay we use the approach of Misner, Thorne and Wheeler [12].  Suppose the photon is moving along a path which is approximated in Cartesian coordinates by $y=b$, $z=0$ for $-a_T\leq x \leq a_R$.  We also assume that $a_T>>b>0$ and $a_R>>b>0$.  Using ${ds}^2=0$, one finds that $dt = \biggl(1+\frac{3M}{2\sqrt{x^2+b^2}}\biggr) \, dx$.  We modify this by replacing it with the corresponding index of refraction (107).  Thus the total time of transit from transmitter to reflector and back is
\begin{equation} t_{TRT} = 2\int_{-a_T}^{a_T} \, \biggl(\; 1+\frac{3M}{2\sqrt{x^2+b^2}}+\frac{\beta_r M}{(x^2+b^2)^2} \; \biggr) \, dx \qquad . \end{equation}
The second and third terms of this integral correspond to the delay effect.  We see that the second term (which when integrated will be called $\Delta \tau$) is 75\% of the general relativity value. The value of $\Delta \tau$ is
\begin{equation} \Delta \tau = 3M\ln \biggl|\frac{(\sqrt{a_R^2+b^2}+a_R)(\sqrt{a_T^2+b^2}+a_T)}{b^2} \biggr| \qquad .  \end{equation}
The third term which will be called $\Delta(\Delta
\tau)$ when integrated has a value \begin{equation}  \Delta(\Delta \tau)\equiv \frac{\beta_r M^2}{b^3} \biggl[\; \arctan\biggl(\frac{x}b\biggr)+\frac{bx}{x^2+b^2}  \; \biggr]\biggl|_{-a_T}^{a_R} \; \approx \frac{\beta_r M^2 \pi}{b^3} \end{equation}
when $a_T$ and $a_R$ are large compared to $b$.  We assume that $\frac{d a_T}{d\tau} \approx \frac{d a_R}{d\tau} \approx 0$ and hence the rate of change of the total time delay is
\begin{equation} \frac{d\;}{d\tau} \biggl( \Delta \tau + \Delta(\Delta \tau) \biggr) \approx \frac{-6M}{b} \biggr( \, 1 + \frac{\beta_r M \pi }{2b^3} \, \biggr) \, \frac{db}{d\tau} \quad .  \end{equation}
Thus, agreement with the general relativity  value would occur if $\frac{\beta_r M \pi}{2b^3}=\frac13$ and hence if $\beta_r=\frac{2b^3}{3\pi M}$.
If $b=r_0 \approx 6.960 \times 10^{10}$ cm, then we find that $\beta_r \approx 4.844 \times 10^{26}$ cm$^2$.  For the stress energy tensor of (57), we find $\alpha_r \approx 9.688\times 10^{26}$ cm$^2$ and for the stress energy tensor of (62), we find $\alpha_r \approx 5.813 \times 10^{26}$ cm$^2$.  We note that these values of $\alpha_r$ are fairly typical. For example, the corresponding value of $\alpha_r$ for hydrogen ($H_2$) gas is approximately $7.9\times 10^{26}$ cm$^2$.  However, the Lorentz-Lorenz relation in its most basic form [18] relates the number density to the refraction.  If the consideration of the corona temperature is correct, the number density of the halo near the sun is approximately $5\times 10^7$ times that of hydrogen gas.  Thus, the dark matter is seen to interact weakly.

As already noted, with $b=r_0$ and $\beta_r=\frac{2b^3}{3\pi M}$, (114) leads to the general relativity result of $\frac{-8M}{b} \frac{db}{d\tau}$ for the time delay.  For the deflection of light problem, we find that $B= \frac{2M}{3\pi r_0}$ and hence (110) yields $\Delta \phi \approx \frac{-4M}{r_0} + \Bigl(\frac{3M}{r_0}+\frac{16M}{3\pi r_0}\Bigr)\sqrt{\frac{3M}{r_0}+\frac{4M}{3\pi r_0}}\;$.  For $M\approx 1.477\times 10^5 $ cm and $r_0 \approx 6.960 \times 10^{10}$ cm, we find that $\Delta \phi \approx -8.462 \times 10^{-6}$.  We note that this value of $\Delta \phi$ is approximately $1.75 ^{\prime\prime}$.

\par \phantom{D} \par

\section{Conclusion.}

The theory based on the conservative transformation group may provide a theoretical basis for a unified field theory and may also provide a theoretical basis for dark matter and the correct modification of general relativity.  It remains to be shown whether the geometry associated with conservative transformations is the correct quantum geometry.  The Lagrangian for the field with sources may be used in a variety of applications, including quantization.  The internal solution and its corresponding stellar model needs additional work to produce more realistic models.  The external solutions, being non-compact, show promise for explaining dark matter.  Excellent agreement is found with Kepler's Law and redshift.  The theory also gives a realistic explanation for the Pioneer anomaly and the high temperature of the corona.  While there are differences in the precession of perihelia, light deflection and time delay predictions, these may be explained by the fact that the stress-energy tensor is non-zero, yielding densities and pressures that affect the motion of planets and photons.

\subsection*{Acknowledgments}  The author would like to thank Dave Pandres for many helpful suggestions.  Also the author would like to thank Peter Musgrave, Denis Pollney and Kayll Lake for the GRTensorII software package which was very helpful.


\end{document}